\documentstyle[namedreferences]{kluwer6}
\def\verbfont{\small\tt}
        
\runningtitle{Particle acceleration and high-frequency
emission in the jets}
\runningauthor{V.V. Usov and M.V. Smolsky}

\begin{opening}

\title{Particle acceleration and high-frequency (X-ray and
$\gamma$-ray) emission in the jets of active galactic nuclei
}
\author{V.V. \surname{Usov} and M.V. \surname{Smolsky}}
\institute{
Department of Condensed Matter Physics, Weizmann Institute of Science, 
Rehovot 76100, Israel} 

\end{opening}

\begin{ao}
Department of Condensed Matter Physics, Weizmann Institute of Science, 
Rehovot 76100, Israel
\end{ao}                   

\begin{document}

\begin{abstract}
It is suggested that the outflowing plasma in the jets of active 
galactic nuclei (AGNs) is inhomogeneous and consists of separate clouds. 
These clouds are strongly magnetized and move away from the central 
engine at relativistic speeds. The clouds interact with an ambient 
medium which is assumed to be at rest. In the process of this 
interaction, particles of the ambient medium are accelerated 
to high energies at the cloud front and flow ahead of the front.
It is shown that the radiation of the 
accelerated particles may be responsible for the X-ray and
$\gamma$-ray emission from AGN jets. TeV $\gamma$-ray emission is 
generated in the inner parts of AGN jets where the Lorentz factor of 
the cloud fronts is $\Gamma_0\geq 30$, while GeV $\gamma$-ray 
emission emanates from the outer parts of AGN jets where $\Gamma_0$ is 
$\sim 10$. 
\end{abstract}


\section{Introduction}

At radio frequencies, where VLBI can resolve the emission regions at
the milliarcsecond scale, many of radio-loud 
AGNs exhibit compact jets (e.g., Cawthorne, 1991).
These jets are remarkably well collimated, with opening 
angles about a few degrees or even less. The radio 
emitting plasma of the jets moves at relativistic velocities. Lower 
limits to the bulk Lorentz factors of this plasma which are derived
from the measured apparent transverse velocities of radio emitting
blobs are between $\sim 3$ and 10 (Impey, 1987).

\par
The Energetic Gamma Ray Experiment Telescope (EGRET) on the
Compton Gamma Ray Observatory (CGRO) has detected a few tens 
of extragalactic sources (Thompson {\it et al.}, 1995),
which are thought to be radio galaxies favorably oriented so that 
the axes of the radio jets are nearly aligned in our observing direction 
(e.g., Dermer, 1994). The $\gamma$-ray emission from 
these EGRET sources which are frequently called blazars
is in the range from a few ten MeV to a few GeV. 
Besides, very hard $\gamma$-rays at TeV energies were detected 
from Mrk 421 and Mrk 501 with the ground-based
telescopes (Punch, 1992; Quinn {\it et al.}, 1996;
Barrau {\it et al.}, 1997). 
Hence, particles are accelerated in AGN jets at least up to 
the energies of $\sim 10^9 -10^{12}$ eV. 

For the jets of AGNs, the typical size of the emission region in 
$\gamma$-rays is somewhere between $\sim 0.01$ pc and $\sim 0.1$ pc 
(e.g., Maraschi, Ghisellini and Celotti, 1992; Levinson, 
1996) that is about one or two orders of magnitude smaller than  
the typical size of the radio emission region. Thus, the $\gamma$-ray
emission most likely emanates from the inner parts of the jet.
The bulk Lorentz factors of the inner $\gamma$-ray jets
may be at least a few times more than the same of the radio 
jets because of deceleration of the jet plasma  in the 
process of its interaction with an ambient medium (see below).

The variability of extragalactic compact jets in the radio and
$\gamma$-ray emission is very high. Most probably, this variability 
at least on some time scales
is due to strong nonstationarity of the plasma flow in AGN jets 
(Blandford and K$\ddot{\rm o}$nigl, 1979; Romanova and Lovelace,
1997; Levinson and van Putten, 1997). In this paper, 
acceleration of particles and their emission in 
relativistic nonstationary jets are discussed.

\section{Acceleration of particles}

Let us consider the following model of the inner high-speed jets 
which are responsible for the $\gamma$-ray emission of AGNs. The 
outflowing plasma of such a jet is inhomogeneous and gathers 
into separate clouds along the jet axis. These clouds are strongly
magnetized and move away from the central engine of AGN at relativistic 
speeds. The Lorentz factor of the cloud fronts is $\Gamma_0\geq 10$.
There is an ambient medium between the clouds which  
is more or less at rest in the frame 
of the AGN engine. This medium mainly consists of electrons and 
protons, and its magnetic field is negligible.

If the jet is free (confined solely by its own inertia), it resembles 
a conic section of a spherical wind. In this case, the components 
of the magnetic field parallel and transverse to the jet velocity are
$B_\parallel \propto r^{-2}$ and $B_\perp\propto r^{-1}$, respectively,
where $r$ is the distance from the central engine. Therefore, at 
large distances from the engine the field of the plasma clouds
is mainly across their velocity, $B_\perp \gg B_\parallel$. 

The plasma clouds in the process of their motion interact with the 
ambient medium at the cloud fronts. For consideration of this 
interaction, it is convenient to switch to the frame
of the cloud front. In this frame, the problem of the 
relativistic magnetized
cloud -- ambient medium interaction is identical to the problem of
collision between a wide relativistic beam of cold plasma and a region
with a strong magnetic field which is called a magnetic barrier.
Recently, such a collision was studied numerically (Smolsky 
and Usov, 1996; Usov and Smolsky, 1998), and it was shown
that when the energy densities of the plasma beam and the magnetic 
field are comparable, 

\begin{equation}
\alpha = 4\pi n_0m_pc^2(\Gamma_0-1)/B^2_0\sim 1\,, 
\label{alpha}
\end{equation}

\noindent
the process of the beam -- barrier interaction is strongly nonstationary, 
and electrons are accelerated up to the mean energy of protons, i.e. up 
to the energy of $\sim m_pc^2\Gamma_0$ (see Fig. 1), where $n_0$ is 
the density of protons and $m_p$ is the proton mass,
$B_0$ is the field strength. Both $n_0$ and $B_0$ are taken in the 
frame of the magnetic barrier. Besides, strong nonstationarity of
the beam -- barrier interaction results in generation of
low-frequency electromagnetic waves in the front vicinity. 
The frequency of these waves is about the gyrofrequency of protons,
$\omega\simeq eB_0/m_pc\Gamma_0$, and their typical amplitude
is $\sim (0.2- 0.3) B_0$.

\begin{figure}
\vspace{7cm}
\caption{Energy spectrum of electrons which are reflected from
the cloud front at $\alpha =2/3$ in the frame of the front.}
\end{figure}

At $\alpha\sim 1$, in the frame of the ambient medium, the mean Lorentz 
factor of high-energy electrons which are accelerated 
at the cloud front and flow ahead of the front is
(Usov and Smolsky, 1998)

\begin{equation}
\langle\Gamma_e\rangle \simeq 0.1
({m_p}/{m_e})\Gamma_0^2
\label{Gammae} 
\end{equation}

\noindent within a factor of 2 or so, where $m_e$ is the electron mass.
The maximum Lorentz factor of accelerated electrons is about an order
of magnitude more than their mean Lorentz factor:

\begin{equation}
\Gamma_e^{\rm max}
\simeq 10 \langle\Gamma_e\rangle\simeq (m_p/m_e)\Gamma_0^2\,.
\label{Gammamax} 
\end{equation}

\noindent
Equations (\ref{Gammae}) and (\ref{Gammamax}) are valid only if the cloud
front velocity is ultrarelativistic, $\Gamma_0\geq 10$.

The energy spectra of accelerated electrons in the range from 
$\sim\langle\Gamma_e\rangle$ to $\Gamma_e^{\rm max}$ depend on 
the value of $\alpha$, and at $0.5\leq \alpha \leq 2$ they
may be fitted by a power law distribution, $dn_e/d\Gamma_e\propto
\Gamma_e^{- \beta}$, with the spectral index $\beta\simeq 2.4-2.8$.

\begin{figure}
\vspace{7cm}
\caption{Energy spectrum of protons which are reflected from
the cloud front at $\alpha =2/3$ in the frame of the front.}
\end{figure}

At $\alpha\sim 1$, the kinetic energy that is lost by the
clouds in the process of their interaction with the ambient 
medium is distributed in the following way (for details, see 
Smolsky and Usov, 1996; Usov and Smolsky, 1998).
About 50\% - 60\% of
this energy is in ultrarelativistic protons that are reflected 
from the cloud front. The rest of the 
kinetic-energy losses is distributed more or less evenly between
high-energy electrons and low-frequency electromagnetic waves that 
are generated in the front vicinity due to nonstationarity of 
the interaction between the relativistic magnetized clouds and 
the ambient medium. Figure 2 shows the energy distribution for reflected 
protons in the frame of the cloud front. In this frame, the mean 
Lorentz factor of reflected protons is about $0.5 \Gamma_0$, while
in the frame of the ambient medium it increases by the factor of
$\sim 2\Gamma_0$ and is
$\langle\Gamma_p\rangle\simeq \Gamma_0^2$.

\section{High-frequency radiation from AGN jets}

\subsection{Gamma-ray emission}

The GeV and TeV $\gamma$-ray emission from AGN jets is
usually interpreted as being produced by inverse Compton scattering of
high-energy electrons on an external radiation which is either the thermal 
UV radiation of the accretion disk around a supermassive black hole 
(e.g., Dermer \& Schlickeiser 1993) or the optical and UV emission 
of warm gas from the broad emission-line region around the AGN 
(e.g., Sikora, Begelman and Rees, 1994; Blandford and Levinson, 
1995; Sikora {\it et al.}, 1996). 

The mean energy of photons after scattering is 
(e.g., Blumenthal and Tucker, 1974)

\begin{equation}
\langle\varepsilon_\gamma\rangle\simeq \varepsilon_0\sin 
\langle\vartheta
\rangle\Gamma_e^2\,,
\label{Egamma} 
\end{equation}

\noindent where $\varepsilon_0$ is the typical energy of external
photons that are scattered by high-energy electrons
with the Lorentz factor $\Gamma_e$, 
$\langle\vartheta\rangle$ is the mean angle between 
the wave vector of external photons and the velocity of
high-energy electrons. Roughly, we have $\varepsilon_0
\simeq 1-10$ eV and $\sin\langle\vartheta\rangle\simeq 1$ for scattering 
of electrons on the radiation of warm gas from 
the broad emission-line region and $\varepsilon_0\simeq 10^2-10^3$ eV
and $\sin\langle\vartheta\rangle\simeq\langle \vartheta\rangle
\simeq 10 r_g/r\simeq 10^{-2}(M_{_{\rm BH}}/10^8 M_\odot )(r/0.01\,
{\rm pc})^{-1}$ for scattering of electrons on the thermal 
radiation of the accretion disk, where $r_g$ is the gravitational radius
of the putative black hole at the center of AGNs, $M_{_{\rm BH}}$ is the 
mass of the black hole which is conventionally about $10^8 M_\odot$. 
We can see that the value of $\varepsilon_0\sin \langle\vartheta\rangle$
is more or less the same for the both kinds of external radiation,
$\varepsilon_0\sin \langle\vartheta\rangle\simeq 1-10$ eV.
From this equation and Equation (\ref{Egamma}), we have
 
\begin{equation}
\langle\varepsilon_\gamma\rangle\simeq (1-10)\Gamma_e^2
\,\,\,{\rm eV}\,.
\label{Egamma1} 
\end{equation}

For $\Gamma_0\simeq 10$, which is the typical Lorentz factor of radio
emitting blobs of AGN jets,
Equation (\ref{Gammae}) yields the mean Lorentz factor of accelerated 
electrons $\langle\Gamma_e\rangle \simeq 10^4$. Substituting 
$\langle\Gamma_e\rangle \simeq 10^4$ for $\Gamma_e$ into Equation
(\ref{Egamma1}), we have the mean energy of photons 
after scattering $\langle\varepsilon_\gamma \rangle
\simeq 0.1-1$ GeV. 

Using Equations (\ref{Gammamax}) and (\ref{Egamma1}),
the maximum energy of photons after scattering is

\begin{equation}
\varepsilon_\gamma^{\rm max}\simeq \min\, [m_ec^2\Gamma_e^{\rm max},
\,10(\Gamma_e^{\rm max})^2\,\,{\rm eV}]\simeq 10^{11}\,\,{\rm eV}\,.
\label{Egammamax} 
\end{equation}

\noindent In Equation (\ref{Egammamax}), $\Gamma_0=10$ is used 
to get the last equality. We can see that for $\Gamma_0\simeq 10$
the expected values of both $\langle\varepsilon_\gamma \rangle$ 
and $\varepsilon_\gamma^{\rm max}$ are consistent with the EGRET data 
on the GeV $\gamma$-ray emission from AGNs. 

As to the TeV $\gamma$-ray emission which was observed from Mrk 421 and
Mrk 501 (Punch, {\it et al.}, 1992; Quinn {\it et al.}, 1996; 
Barrau {\it et al.}, 1997), it may be explained by inverse 
Compton scattering of high-energy electrons which are accelerated at the 
front of a relativistic  magnetized cloud only if $\varepsilon_\gamma^
{\rm max}\geq 10^{12}$ eV. From this condition and Equations 
(\ref{Gammamax}) 
and (\ref{Egammamax}), we have the lower limit on the Lorentz factor 
of the cloud fronts, $\Gamma_0\geq 30$. The inverse Compton scattering
producing the TeV $\gamma$-ray emission from both Mrk 421 and Mrk 501
will be in the Klein-Nishina limit where the scattered photon energy
is comparable to the electron energy. In this case, the differential 
spectral index of photons after scattering coincides with the
differential spectral index of high-energy electrons. The 
spectra of Mrk 421 and Mrk 501 in TeV $\gamma$-rays are not
significantly different, and their differential spectral indexes
are $\sim 2.5$ (e.g., McEnery {\it et al.}, 1997,
Quinn {\it et al.}, 1997;
Zweerink {\it et al.}, 1997) that is consistent with 
the differential spectral index of high-energy electrons accelerated
in the cloud front vicinity (see Section 2).

The high-energy protons which are accelerated in the process of their
reflection from the cloud front may be a powerful source of TeV 
$\gamma$-rays too (cf. Mannheim and Biermann, 1992; Dar and Laor,
1997; Bednarek and Protheroe, 1997). 
TeV $\gamma$-rays may be produced efficiently by the 
interaction of the high-energy protons in the AGN jet with the
protons of the ambient medium via, for example, $pp\rightarrow \pi^0X$; 
$\pi^0\rightarrow 2\gamma$ (Dar and Laor, 1997). In this case,
the mean energy of generated $\gamma$-rays is 

\begin{equation}
\langle\varepsilon_\gamma
\rangle\simeq 70\langle\Gamma_p\rangle\,\,\,{\rm MeV}\,. 
\label{meg} 
\end{equation}

Taking into account that the mean Lorentz factor of protons reflected from  
the wind front is $\langle\Gamma_p\rangle\simeq \Gamma_0^2$,
from Equation (\ref{meg}) it follows that the bulk
photons is in the TeV range, $\langle\varepsilon_\gamma\rangle
\geq 10^{12}$ eV, only if the Lorentz factor of the cloud front is
extremely high, $\Gamma_0\geq 10^2$.

\subsection{ X-ray emission}

High-energy electrons which are accelerated at the cloud front 
and move ahead of the front interact not only with high-frequency 
(optical, UV and X-ray) photons, but with the low-frequency 
electromagnetic waves too (about these waves, see Section 2). 
The motion of electrons and their radiation in the fields of
electromagnetic waves is characterized by 
the following dimentionless Lorentz-invariant parameter
(e.g., Blumenthal and Tucker, 1974):

\begin{equation}
\eta ={e\tilde B\over mc\omega}\,,
\label{eta}
\end{equation}

\noindent
where $\tilde B$ is the wave amplitude and $\omega$ is the wave 
frequency. At $\eta\ll 1$, electrons radiate via Compton 
scattering. In this case, the energy spectrum of photons after
scattering does not dependent on the wave
amplitude (see Equation (\ref{Egamma})). 
Radiation of electrons in the fields of strong electromagnetic 
waves, $\eta\gg 1$, is called as Compton-synchrotron radiation.
This radiation closely resembles synchrotron radiation.
Indeed, it is well known that electromagnetic radiation of 
relativistic electrons is concentrated in the direction
of the particle's velocity within a narrow cone of angle 
$\Delta \varphi\simeq 1/\Gamma_e \ll 1$, and this  
radiation may be observed only if the observer is inside of the
cone (e.g., Rybicki and Lightman, 1979). The formation length of  
radiation of relativistic electrons in a magnetic field $B$
is about $\Delta l \simeq 2 R_{\rm B}\Delta \varphi\simeq 2 c/
\omega_{\rm B}=2m_ec^2/eB$, where $R_{\rm B}=(c/\omega_{\rm B})
\Gamma_e$ is the electron gyroradius and
$\omega_{\rm B}=eB/m_ec$ is the electron gyrofrequency.
For the Compton-synchrotron radiation, $\eta\gg 1$,
the formation length is much smaller than the 
wave-length, $\lambda$, of the strong electromagnetic waves, 
$\Delta l\simeq (\omega/\pi \omega_{\rm B})\lambda =\lambda /
\pi\eta \ll \lambda$, and the electromagnetic fields of the strong 
waves may be considered as infinite and static, where  $\lambda 
=2\pi c/\omega$. Therefore, the Compton-synchrotron radiation  
is like the synchrotron 
radiation in the magnetic field which is equal to the local
magnetic field of the waves. For example, the mean energy of photons
which are generated via
the Compton-synchrotron radiation is $\nu_{sc}\simeq (\omega_{\rm B}
/2\pi )\Gamma_e^2$ that coincides with the mean energy of 
synchrotron photons and qualitatively differs from the mean energy of
photons after Compton scattering (cf. Equation (\ref{Egamma})).
[For details on Compton-synchrotron 
radiation see (Gunn and Ostriker, 1971; Blumenthal 
and Tucker, 1974).] To avoid a misunderstanding, it is worth
noting that the so-called synchrotron-self-Compton (SSC) models are 
frequently discussed for blazars (e.g., Maraschi {\it et al.},
1992). In these models, the soft photons are arised from the
synchrotron emission, and then, $\gamma$-rays are generated via
Compton scattering of these soft photons.

For the low-frequency waves with $\tilde B
=\tilde B_{_{\rm LF}}\simeq (0.2-0.3)B$ 
and $\omega= \omega_{_{\rm LF}}\simeq eB/m_pc \Gamma_0$
which are generated at the cloud front, we have

\begin{equation}
\eta \simeq {\tilde B_{_{\rm LF}}\over B}{m_p\over m_e}\Gamma_0
\gg 1\,,
\label{eta1}
\end{equation}

\noindent
where $B\simeq \Gamma_0 B_0$ is the magnetic field of the cloud
in the frame of the ambient medium, i.e.
in the frame of the central engime of AGN. 
In this case, the typical frequency of Compton-synchrotron radiation 
of high-energy electrons is (e.g., Blumenthal and Tucker, 1974)

\begin{equation}
\nu_{sc}\simeq {e\langle\tilde B_{_{\rm LF}}\rangle \Gamma^2_e
\over 2\pi mc}\simeq 3\times 10^6\langle\tilde B_{_{\rm LF}}\rangle
\Gamma_e^2\,\,\,\,{\rm Hz}\,,
\label{nu}
\end{equation}

\noindent
where $\langle\tilde B_{_{\rm LF}}\rangle\simeq {1\over 2} 
\tilde B_{_{\rm LF}}\simeq 0.1 B$  is the mean field value of 
strong low-frequency waves. 

For a Poynting flux-dominated jet, the strength of the cloud magnetic 
field at the distance $r$ from the central engine is 

\begin{equation}
B\simeq 10^2\left(
{L\over 10^{47}\,{\rm ergs\,\,s}^{-1}}\right)^{1/2}
\left({r\over 0.01\,{\rm pc}}\right)^{-1}\,\,\,{\rm G}\,,
\label{B}
\end{equation}

\noindent where $L=4\pi r^2c(B^2/8\pi )$ is the total luminosity  
for a spherical Poynting flux-dominated wind. 
Such a jet may be roughly considered as a conic section of this wind. 

From Equations (\ref{Egamma}), (\ref{nu}) and (\ref{B}), the ratio
of the mean energy of Compton-synchrotron photons and
the mean energy of $\gamma$-rays is

\begin{equation}
{\langle\varepsilon_{sc}\rangle \over \langle\varepsilon_\gamma\rangle}
\simeq 2\times 10^{-8}
\left({\varepsilon_0\sin \langle\vartheta\rangle\over 1\,\,{\rm eV}}
\right)^{-1}\left(
{L\over 10^{47}\,{\rm ergs\,\,s}^{-1}}\right)^{1/2}
\left({r\over 0.01\,{\rm pc}}\right)^{-1}.
\label{epsilons}
\end{equation}

In the case of Mrk 421 for its typical parameters,  
$L\sim 10^{47}$ ergs s$^{-1}$, $r\sim 0.01$ pc, 
$\varepsilon_0\sin \langle\vartheta\rangle\sim 3$ eV and
$\langle\varepsilon_\gamma\rangle\sim 10^{12}$ eV,
Equation (\ref{epsilons}) yields $\langle\varepsilon_{sc}\rangle
\sim 10$ keV that is consistent with the 
multifrequency observations of Mrk 421 (Macomb {\it et al.}, 1995;
Schubnell {\it et al.}, 1997; Zweerink {et al.}, 1997). Fits
of X-ray and $\gamma$-ray spectra of Mrk 421 and another 
extragalactic sources in the frame of presented model 
are under way.

\section{Conclusion and discussion}

We have considered in this paper the generation of X-ray and 
$\gamma$-ray emission from the AGN jets. High-energy electrons 
which are responsible for this emission are
accelerated inside the jets in the process
of interaction between relativistic magnetized clouds and an
ambient medium which is assumed to be nearly at rest. For a powerful
extragalactic source of TeV $\gamma$-rays like Mrk 421, the Lorentz 
factor of these clouds is $\Gamma_0 \geq 30$.
Recently, it was shown (Renaud and Henri, 1997) that
in the case of supermassive black holes relevant to AGNs 
the terminal Lorentz factor of the jet plasma  
may be as high as 60 that is more than the lower limit on $\Gamma_0$.

For high-energy electrons which are accelerated at the cloud front,
the characteristic time of their energy losses is

\begin{equation}
\tau _{loss}\simeq {5\times 10^8\over (\langle\tilde 
B_{_{\rm LF}}^2\rangle +4\pi u_{ph})\Gamma_e}\,\,\, {\rm s}\,,
\label{Tau}
\end{equation}

\noindent
where $u_{ph}$ is the energy density of soft photons in
ergs cm$^{-3}$ and $\langle\tilde B_{_{\rm LF}}^2\rangle$ in G$^2$. 
For typical parameters, $\langle\tilde B_{_{\rm LF}}^2\rangle\simeq 
4\pi u_{ph}\simeq 10$, from Equations 
(\ref{Gammae}) and (\ref{Tau}) we have $\tau_{loss}$ is $\sim 
2\times 10^3$ s at $\Gamma_0=10$ and $\sim 2\times 10^2$ s at 
$\Gamma_0=30$. We can see that the value of $c\tau _{loss}$ is much 
smaller than the typical size of the emission region in $\gamma$-rays
which is $\sim 10^{17}$ cm (Maraschi {et al.}, 1992; 
Levinson, 1996). Therefore, high-energy electrons do not propagate
far from the barrier front because of thier energy losses.
As to propation of the high-energy protons through the
ambient medium, it depends essentially on the magnetic field of
the ambient medium. Consideration of interaction between the protons  
which are reflected from the cloud front and the ambient medium
with taking into account development of two-stream instability is 
under way and will be addressed elsewhere. 

It is worth noting that the Compton scattering of photons which are
generated via the Compton-synchrotron emission may be a powerful
source of $\gamma$-rays from the AGN jets as well. 
However, we did not consider this process
because, in fact, it is a version of
the synchrotron-self-Compton models in which the Compton-synchrotron
radiation is used instead of the synchrotron radiation. 

Not any kind of the ambient 
medium is suitable for strong acceleration of electrons at 
the cloud front. For example, if the ambient 
medium consists of $e^+e^-$ pairs, the discussed mechanism of  
electron acceleration does not work, and 
Equations (\ref{Gammae}) and (\ref{Gammamax}) are out of their 
applicability (Smolsky and Usov, 1996).

As noted above, electrons of the ambient medium may be strongly 
accelerated in the cloud front vicinity only 
if $\alpha$ is $\sim 1$ (Usov and Smolsky, 1998). At 
$\alpha \sim 1$, from Equation (\ref{alpha}) the density of the 
ambient medium in its frame is of the order of

\begin{equation}
n_1 = {B^2\over 4\pi \Gamma_0^4m_pc^2}\,.
\label{n1}
\end{equation}

\noindent
For $r\simeq 0.01$ pc, $L\simeq 10^{47}$ ergs s$^{-1}$ and 
$\Gamma_0\simeq 30$, from Equations (\ref{B}) and (\ref{n1}) 
we have $B\simeq 10^2$ G and $n_1\simeq 1$ cm$^{-3}$.
 
There are a few reasonable sources of the ambient medium in 
the AGN jets. They are compact clouds of warm gas, stellar winds or
a trail of plasma which is left after ejection of a
relativistic magnetized cloud 
(on interaction between the AGN jets and compact clouds of 
warm gas, see Higgins, O'Brein and Dunlop, 1995;
Bicknell, Dopita and O'Dea, 1997). It is difficult to say now
about what is the main source of the ambient medium. 
All these sources are able,
in principle, to be responsible for the ambient medium with such a
low density as $n_1\simeq 1$ cm$^{-3}$ (about stellar winds
as a source of the ambient medium, see below).

At the central regions of AGNs, where the $\gamma$-ray 
emission is generated, the density of stars is very high, and the mean 
stellar density may be as high as $\sim 10^8\,M_\odot\,{\rm pc}^{-3}$ 
or even more (e.g., Lauer {\it et al.}, 1992). Massive OB stars, 
$M_{_{\rm OB}}\sim 10M_\odot$, contain a substantial part 
of the mass of the stellar clusters at the AGN center. Therefore, it is
natural to expect that there is at least a few such massive stars at the
distance $r\leq 0.01$ pc from the AGN engine. 

Massive stars are characterized by intense winds (e.g., Garmany and Conti, 
1984; Leitherer, 1988). For a typical massive OB star, the 
mass-loss rate is $\dot M_{_{\rm OB}}\sim 10^{-7}-
10^{-6}M_\odot\,{\rm yr}^{-1}$, 
and the terminal velocity of the matter outflow is $v_\infty\simeq
(1-3)\times 10^8$ cm s$^{-1}$. 

The wind density at the distance $r_w$ from the OB star is 

\begin{equation}
n_w\simeq {\dot M_{_{\rm OB}} \over 4\pi r_w^2v_\infty m_p}\,.
\label{nw}
\end{equation}

\noindent
For $\dot M_{_{\rm OB}}\simeq 10^{-7}M_\odot$ yr$^{-1}$, $v_\infty
\simeq 2\times 10^8$ cm s$^{-1}$, $r_w\simeq 0.01$ pc, from Equation 
(\ref{nw}) we have $n_w\simeq 1$ cm$^{-3}$. Hence, the stellar 
winds from massive stars may be one of the reasonable sources of 
the ambient plasma with the density $n_w\sim n_1$ which is
favorable for strong acceleration of electrons in the AGN jets.

A remarkable change in the TeV $\gamma$-ray emission of Mrk 501 
was seen in 1997: the flux of TeV $\gamma$-rays increased dramatically 
from previous seasons (e.g., Quinn {\it et al.}, 1997). In April 
and May, 1997, Mrk 501 at times was the brightest TeV source in the sky. 
The average emission level has increased by a factor of more than 16 since 
its discovery as a TeV $\gamma$-ray source in 1995. Maybe, this long-time 
variability of Mrk 501 in TeV $\gamma$-rays results from the orbital
motion of a OB star around the putative supermassive black hole
at the center of Mrk 501. Indeed, in the process of orbital motion of
a massive OB star around the black hole the ambient-plasma density
in the jet may vary with the orbital period, $P$,
of the star:

\begin{equation}
P\simeq 10
\left({M_{_{\rm BH}}\over 10^8\,M_\odot}\right)^{-1/2}
\left({r\over 0.01\,{\rm pc}}\right)^{3/2}\,\,{\rm yr}\,,
\label{P}
\end{equation}

\noindent
and the most favorable condition, $\alpha\simeq 1$ or
$n_w\simeq n_1$, for acceleration of electrons in the AGN jet may occur
at times. As it is mentioned in Section 2, in this case, when
$\alpha$ is $\sim 1$, the luminosity of the jet 
in $\gamma$-rays is extremely high. Maybe, such a brightening of 
Mrk 501 in TeV $\gamma$-rays was observed in 1997. If this suggestion
is right, it is expected that the TeV $\gamma$-ray emission of Mrk 501 
is modulated with the period $P\simeq 10$ yr,
and the duration of the stage with a strong TeV $\gamma$-ray emission is 
about an order of magnetude smaller than $P$.

\begin{acknowledgements}
We thank M. Milgrom for helpful conversations.
This research was supported in part by MINERVA Foundation,
Munich / Germany. 
\end{acknowledgements}

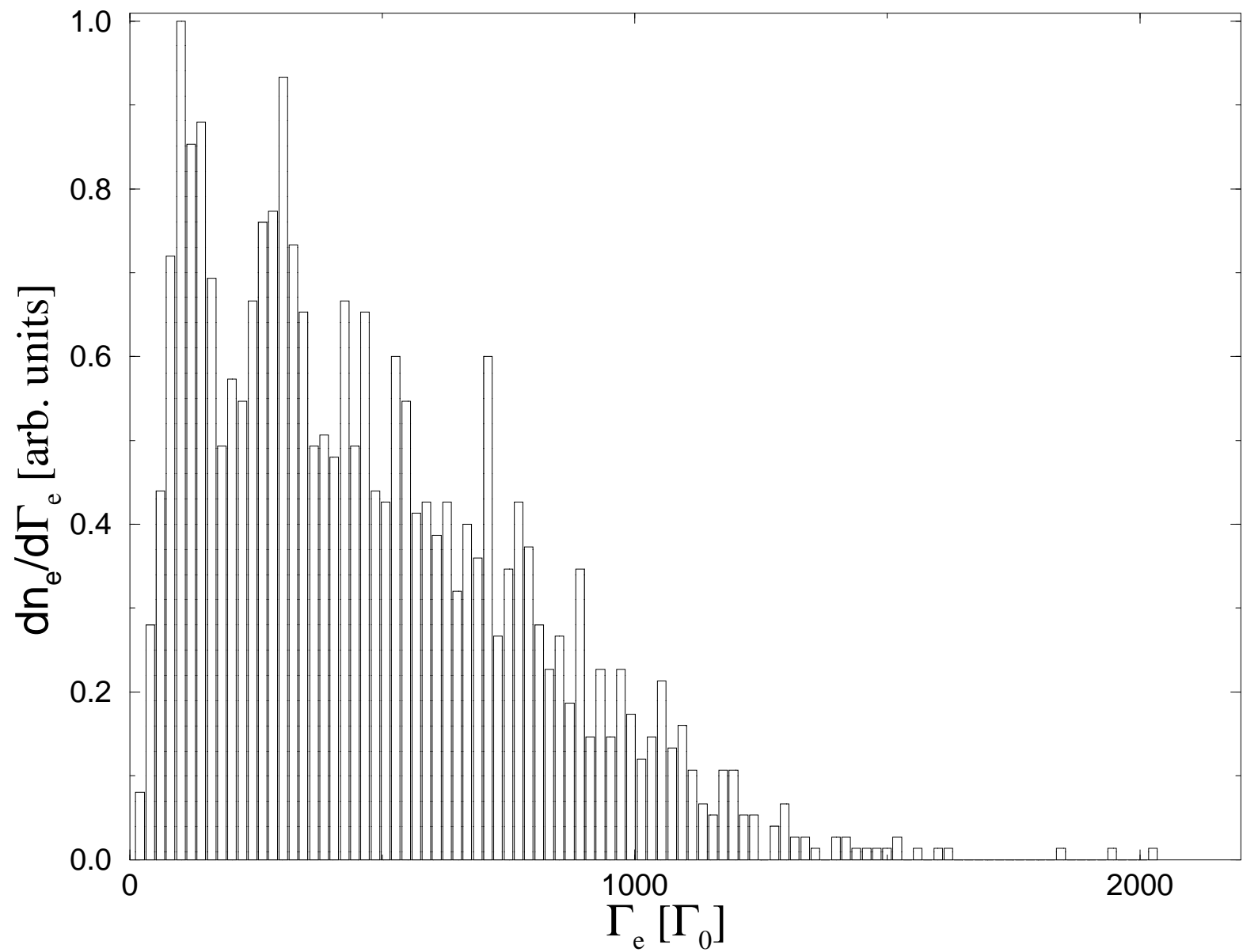
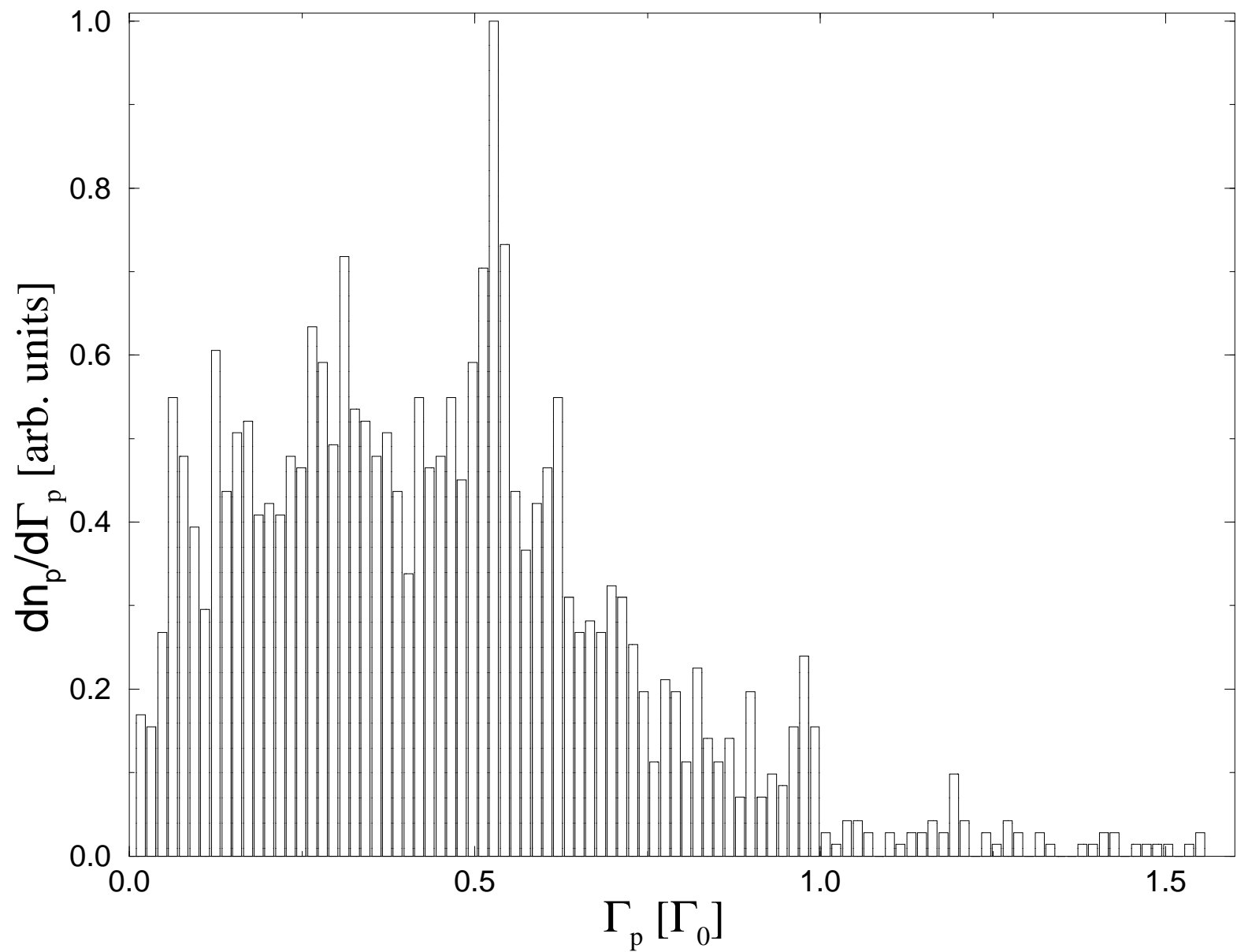
\end{document}